\documentclass[aps,prb,twocolumn,showpacs,floatfix,superscriptaddress]{revtex4}
\usepackage{graphicx}
\usepackage{multirow}
\usepackage{mathtools}
\usepackage{amssymb}
\usepackage{amsfonts}
\usepackage{mathrsfs}
\usepackage{color}
\usepackage[normalem]{ulem}

\newcommand{\vect}[1]{\boldsymbol{#1}} 

\begin{document}
\title{Hedin equations in resonant microcavities}
\author{Paolo E. Trevisanutto}\affiliation{Centre for Advanced 2D Materials and Graphene Research Centre, National University of Singapore, Singapore, 117546}
\affiliation{Singapore Synchrotron Light Source, National University of Singapore, 5 Research Link, Singapore 117603, Singapore}
\author{Mirco Milletar\`{i}}\affiliation{Centre for Advanced 2D Materials and Graphene Research Centre, National University of Singapore, Singapore, 117546}
\date{\today}
%
%
\begin{abstract}
With the improvement of the experimental techniques, many new phenomena in which the photon degrees of freedom are involved have been discovered. Typical examples are the exciton-polariton quasi-particles, excitons strongly coupled to photons. A correct description of these systems requires a full account of the photon's dynamics, including the vector gauge degrees of freedom. In order to include this contribution in \emph{ab initio} many-body perturbation theories, here we present a generalization of the Hedin equations, originally derived for a many electron system with Coulomb interaction. These equations are now derived to take into account the strong coupling physics of the electrons and the \textit{transverse} photon's degrees of freedom.
\end{abstract}


\maketitle

\section{Introduction}
In the past decade, progress in designing semiconductor microcavities has attracted much interest from different fields of physics, ranging from condensed matter to quantum optics and relativistic particle physics~\cite{Exciton-polariton_np,Carusotto,Keeling}. Typically, a microcavity is a system where a two dimensional (2D) quantum well (QW) is embedded in Bragg mirror barriers. The confinement of the photons enhances the external electromagnetic (EM) interaction with the QW so that retardation effects due to the EM vector potential cannot be considered negligible, even though the system is in a non relativistic regime. When strongly coupled to the photons, elementary boson excitations can give rise to new quasiparticles: \textit{polaritons}. Typical examples are the\textit{ exciton-polaritons}, photons dressed by electron-hole excitations bearing a very light effective mass ($m_{e} \sim 10^{-4}$).  
Other polaritons stemming from the ultra-strong coupling regime~\cite{Delteil} were recently the subject of intensive investigations due to the possibility of tailoring their properties by tuning the size of the QW, and its potential use for laser polariton devices~\cite{Colombelli,Zanotto}. The ultra-strong coupling regime has been observed in doped semiconductor QWs, where plasmons originated in the interconduction subband result in stable and robust intersubband (ISB) polaritons \cite{Dini_2003,Ciuti_Bastard_Carusotto,Todorov,Geiser,Delteil}. A similar effect takes place in the presence of an applied magnetic field, where the photon mode is coupled to 2D electron gas magneto-plasmon excitations~\cite{Scalari}. Since polaritons are themselves bosons, Bose-Einstein Condensation (BEC) phenomena can be observed and investigated~\cite{Exciton-polariton_np}.

From a theoretical point of view, microcavity quantum electrodynamics (QED) systems are usually described by analytical models \cite{Hopfield, Ciuti_Bastard_Carusotto,Todorov,Todorov_prb,DeLiberato_prl, DeLiberato}, whereas first-principle approaches have been proposed only recently, extending the time-dependent density-functional theory (TD-DFT)
to the relativistic regime~\cite{tddft-qed2}. Any concrete applications of TD-DFT methods still require an appropriate approximation scheme for the electron-photon exchange-correlation (xc) functionals~\cite{tddft-qed}.

Recently, an xc potential for such systems has been proposed within the time dependent optimized effective potential theory (TD-OEP) framework \cite{tddft-oep}, enabling the possibility of first-principles calculations. In contrast, \textit{ab initio} many body perturbation theory (MBPT) approaches \cite{Onida}, which together with TD-DFT, have been very successful in describing the excited states of condensed matter systems, have not been used so far to study the microcavity problem. \textit{Ab initio} MBPT calculations are based on the self consistent solution of the Hedin equations, originally developed for studying electron interactions meditated by the instantaneous Coulomb interaction. These equations were then generalized to take into account spin-orbit and spin-spin interactions~\cite{Arya_bierm} or to be implemented within the dynamical mean field theory framework (DMFT)~\cite{Held}. \\
In this paper, we present the generalization of the Hedin equations to non relativistic systems of charged particles, where the quantum degrees of freedom of the \textit{transverse} EM \textit{gauge} field is taken into account. The new Hedin equations, which are solved self-consistently, are the basis of possible \textit{ab initio}  MBPT calculations, allowing to obtain the electron and photon quasi particle self energy.

In section \ref{derivation}, we introduce the interactions in the system following the gauge principle~\cite{zee, Chang, Starke}, as done in QED. From the gauged Lagrangian, and in order to make connection with the condensed matter literature, we then obtain an effective Hamiltonian solely defined in terms of the fermionic degrees of freedom. In the second part, we use the effective fermionic description to formulate the problem in terms of the Schwinger functional approach discussed in Ref.~\cite{Strinati}. Finally in section \ref{polariton_sec}, we apply our formalism to the historical case of the exciton-polariton in microcavity QED. Conclusions are drawn in section \ref{conclusion}. The formal derivation of the Hedin equation is presented in the appendix.
\section{Theoretical derivation}\label{derivation}
\begin{figure}[t!]
\includegraphics[width=\columnwidth]{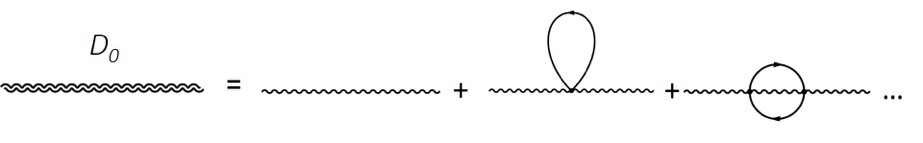}
\caption{\label{Fig1} {\bf Diagrams contributing to the photon propagator}: the first terms in the diagrammatic expansion of the renormalized photon propagator, Eq. \eqref{phprop}. }
\end{figure}

The Lagrangian density for a system of non relativistic electrons is our starting point; by demanding invariance with respect to $U(1)$ gauge transformations this reads
\begin{align} \label{lagr}
\mathscr{L} &= \psi^{\dagger} \left\{ \imath (\partial_t-\imath \, e \, a_0)+ \frac{1}{2 m} (\nabla_i - \imath \, e \, A_i)^2 \right\} \psi \\ \nonumber
&- \frac{1}{2} a^{0} (\mathscr{D}^{-1}_{(0)})_{00} a^{0}- \frac{1}{2} A^{i} (\mathscr{D}^{-1}_{(0)})_{ij} A^{j}.
\end{align}
Here $\psi(1)$ is a fermionic (Grassman) field, $a_0(x)$ is the internal scalar potential (due to electron-electron interactions) and $A_{i}$ is the external (laser) EM gauge field.  We define the total EM gauge field as $A_{\mu}(x)= (c \, a_0(x), A_i(x))$ and the bare gauge field propagator as $(\mathscr{D}_{(0)})_{\mu \nu}= (\eta_{\mu \nu} \partial^2 - \partial^{\mu} \partial^{\nu})$~\cite{Valerio_archiv, zee}. We have also introduced the flat metric tensor $\eta_{\mu \nu}$ and the index $1\equiv (\vect{x}_1,t_1,s)$ stands for space, time and spin variables. In the following we will use Einstein's convention for summation over dummy indices and $\mu = (0, i)$, $i=x,y,z$ (the temporal and spatial components, respectively) and units of $\hslash=1$. In our reference frame, $x$ and $y$ are the \textit{transverse} direction and $z$ is the \textit{longitudinal} direction. Nevertheless, as we focus on non-relativistic systems, the covariant and contravariant indices are indistinguishable. Choosing the Coulomb gauge $\vect{\nabla} \cdot \vect{A} =0$, the propagator of the gauge fields is diagonal, and in momentum space it reads~\cite{Starke}
\begin{equation}
(\mathscr{D}_{(0)})_{\mu \nu}(\omega, \vect{k})= \left(\begin{array}{cc} \frac{4\pi}{|\vect{k}|^2} & 0 \\0 & \frac{4\pi}{\omega^2-c^2 |\vect{k}|^2} \left( \eta_{ij}- \frac{k_i k_j}{|\vect{k}|^2}\right) \end{array}\right),
\label{BarePhoton_prop}
\end{equation}
where one can recognize the Coulomb propagator in the top left corner. Equation ~\eqref{lagr} can now be written in the equivalent form
\begin{align} \label{lagr1}
\mathscr{L} &= \psi^{\dagger} \left\{ \imath \partial_t+ \frac{\vect{\nabla}^2}{2 m}  \right\} \psi + \rho \, a_0 + j^i A_i
\\ \nonumber
&- \left(\frac{e}{ 2m} \right) \rho \, A_{i}A_{i} - \frac{1}{2} A^{\mu} (\mathscr{D}^{-1}_{(0)})_{\mu \nu} A^{\nu},
\end{align}
where $\rho(x)\equiv j_{0}(x)/c =e \, \psi^{\dagger}(x) \psi(x)$ is the charge density and $j_{i}(x)=1/( 2m \imath)  \left[\psi^{\dagger}(x)\left(\nabla_{i}  \psi(x) \right)-\left(\nabla_{i} \psi^{\dagger} (x) \right) \psi (x)\right]$ is the (paramagnetic) charge current. Note that contrary to the relativistic theory, a term proportional to $\vect{A}^2$ (diamagnetic term) now appears. The effect of this term is to renormalize the spatial component of the gauge field propagator. It is then convenient to define the renormalized gauge field propagator as
\begin{equation}\label{phprop}
(D^{-1}_{(0)})_{\mu \nu}= (\mathscr{D}^{-1}_{(0)})_{\mu \nu}+ \frac{e}{m } \rho \, \eta_{\mu,i} \, \eta_{\nu,j},
\end{equation}
where the propagator should be understood in terms of its diagrammatic expansion  (see Fig.~\eqref{Fig1}).
Integrating out the gauge fields, we arrive at the effective Hamiltonian density containing only the fermionic degrees of freedom,
%
%
\begin{equation}
\begin{split}
\mathcal{H}_{MB}(1)= & \mathcal{H}^{0}(1)+\hat{j}^{\mu}(1) \, {\zeta}_{\mu}(1) \\
&-\frac{1}{2} \int d2 \,  \hat{j}^{\mu}(1)\, ({D}_{(0)})_{\mu \nu}(1,2) \, \hat{j}^{\nu} (2),
\end{split}
\label{MB-Hamy}
\end{equation}
%
where $\mathcal{H}^{0}$ is the non-interacting Hamiltonian density
\begin{align} \label{eq:h0}
\begin{split}
\mathcal{H}^{0}(1)&=\hat{\psi}^{\dagger}(1) \left\{ -\frac{\vect{\nabla}^{2}}{2m} +V(\vect{r}_1) \right\} \hat{\psi}(1) \\
& \equiv\hat{\psi}^{\dagger}(1) \, h(1) \, \hat{\psi}(1).
\end{split}
\end{align}
%
In Eqs.~\eqref{MB-Hamy} and ~\eqref{eq:h0} we have introduced the crystalline nuclear potential $V$ and the source field $\zeta_{\mu}$ needed to generate current expectation values. Finally, note that we have moved to the operator formalism and we will designate operator valued quantities with a hat.
In the effective description, the absence of the gauge field is traded for the non local current-current interaction term. If the spatial components of the current are neglected, the well known result for the instantaneous Coulomb interaction in terms of a non local density-density interaction term is recovered ~\cite{zee, Chang, Valerio_archiv}.

The one and two point electron Green's functions are defined in the same way as the time ordered expectation value,
\begin{align} \label{Gfunct}
G_1(1,2) &= \left\langle N\left|  T\left[\hat{ \psi}(1)\hat{ \psi}^{\dagger}(2)\right]\right| N  \right\rangle \\ \nonumber
G_2(1,2;1',2') &= \left\langle N\left|  T\left[\hat{ \psi}(1)\hat{ \psi}(2)\hat{ \psi}^{\dagger}(2')\hat{ \psi}^{\dagger}(1')\right]\right| N  \right\rangle,
\end{align}
where $|N \rangle$ is the many body ground state. Given the above definitions, the charge and current densities can be expressed as
\begin{align}
\rho(1)&=-\imath \, e \, G_1(1,1^+) \\ \nonumber
j_{i}(1)&= -\left( \frac{\hat{\nabla}_{i}-\hat{\nabla}'_{i}}{2m}\right) _{1'\to 1^+} G_1(1,1')= \hat{\xi}_{i}(1,1') G_1(1,1'),
\end{align}
%
%
%
where $1^{+}$ stands for $t_1+\varepsilon$ $(\varepsilon \to 0^+)$. In order to be consistent with this notation $\hat{\xi}_{0}(1,1')\equiv c \, \delta(1,1')$. From the Hamiltonian density of Eq.~\eqref{MB-Hamy}, the equation of motion (EOM) for the electron Green's functions is obtained in analogy to those in Ref.~\cite{Strinati}:
\begin{widetext}
\begin{equation}
 \begin{split}
 &\left(i\frac{\partial}{\partial t_1} -h(1)\right) G_1(1,2)-\int d3\left( G_1(1,3) \zeta_{0}(3)\delta(3,2)+ \frac{\nabla^{i}_{1}}{2m} G_1(1,3) \zeta_{i}(3)\delta(3,2) \right) \\
 &	+ i\int d3 \left(({D}_{(0)})_{00} (1,3)G_{2}(1,3^{+};2,3^{++}) + ({D}_{(0)})_{ij} (1,3) \frac{{\nabla}^{i}_{1}}{2m} \hat{\xi}_{j}(3^{+},3^{+}) G_{2}(1,3^{+};2,3^{++}) \right) =\delta(1,2).
 \end{split}
\label{Heisenberg_eq_mot} 	
\end{equation}
\end{widetext}
%
\begin{table} [t!]
\caption{ The Hedin equation in the presence of the external \textit{transverse} EM field}
\centering
\begin{ruledtabular}
\begin{tabular}{c}
Summary table of the Hedin equations \\
 \hline
 {$G_{1} = G^{(0)}_{1}+G^{(0)}_{1} \, \Sigma \, G_{1}$}\\[2mm]
 {$D_{\mu\nu} = (D_{(0)})_{\mu\nu}  + (D_{(0)})_{\mu\eta} \Pi_{\eta\theta} D_{\theta\nu}$} \\[4mm]
 {$\Sigma = -\Gamma^{(0)\mu} D_{\mu\nu} G_1 \Gamma^{\nu}$}  \\[2mm]
 {$ \Gamma^{\mu} = \Gamma^{(0)\mu}+ \frac{\delta  \Sigma}{\delta G_1} G_1G_1 \Gamma^{\mu}$}\\[2mm]
 {$ \Pi_{\mu\nu} = \Gamma^{(0)\mu} \, G_1G_1 \Gamma^{\nu}$}\\[2mm]
\end{tabular}
\end{ruledtabular}
\end{table}
%
%
This EOM is comprised of the zero-component part describing the usual longitudinal scalar part $({D}_{(0)})_{00}$ (the Coulomb potential) and the new \textit{transverse} components due to the EM gauge field. The formal derivation of Hedin equations is shown in the Appendix. Here we report the final results:
The Dyson equation for the one-particle electron Green's function has the usual form:
\begin{equation}
G_1(1,2)= G^{(0)}_1(1,2)+ \int d(34) \, G^{(0)}_1(1,3) \, \Sigma(3,4) \, G_1(4,2),
\label{Green_Dyson}
\end{equation}
where $\Sigma$(1,2) is the self energy operator and $G^{(0)}_1$ is the Green's function for the non-interacting Hamiltonian.
The  Dyson equation for the self-energy $\Sigma$ is
%
%
\begin{equation} \label{SE_Dyson}
\begin{split}
&\Sigma (1,2)\equiv \\
& -\imath\int \, d(3456) \, \Gamma^{(0)\mu}(6,5;1) D_{\mu\nu} (6,3)  G_1(5,4) \, \Gamma^{\nu}(4,2;3).
\end{split}
\end{equation}

Note that in these equations, we have introduced the \textit{dressed} Photon propagator $D _{\mu\nu}$ (as before  $D_{00}$ is the screened potential usually written as $W$) and $\Gamma^{\nu}$ is the \textit{irreducible} vertex operator .
Therefore, the "GW" self energy is now retrieved by replacing the vertex $\Gamma^{\nu}$ with the bare one, defined as
%
%
\begin{equation}
\Gamma^{(0)\mu}(1,2;3)\equiv\hat{\xi}_{\mu}(3,3')\delta(1,3')\delta(2,3).
\label{Bare_vertex}
\end{equation}
%
%
The Dyson equation for the \textit{irreducible} vertex operator is
\begin{equation}
\begin{split}
\Gamma^{\mu}(1,2;3)=&
\Gamma^{(0)\mu}(1,2;3)+\\
  &\int d(4567) \, \frac{\delta \Sigma(1,2)}{\delta G_1(4,5)} G_1(4,6)G_1(7,5) \Gamma^{\mu}(6,7;3).
\end{split}
\label{Vertex_Dyson}
\end{equation}
%
The Dyson equation of the screened photon Green's function is
\begin{equation}
\begin{split}
D_{\mu\nu} (1,2)=& (D_{(0)})_{\mu\nu} (1,2) +\\
 &\int d(34) \, (D_{(0)})_{\mu\eta} (1,3) \, \Pi_{\eta\theta}(3,4)\, D_{\theta\nu}(4,2),
\label{Screened_dy}
\end{split}
\end{equation}
%
%
%
%
whereas the Dyson equation for the polarization function is
\begin{equation}
\begin{split}
&\Pi_{\mu\nu}(1,2)=\\
&\int d(4567) \, \Gamma^{(0)\mu}(7,6;1) \, G_1(6,4) \, G_1(5,7) \, \Gamma^{\nu}(4,5;2).
\end{split}
\label{Polar_Dyson}
\end{equation}
The set of Eqs.~\eqref{Green_Dyson}-\eqref{Polar_Dyson} (summarized in Table. 1) constitutes the non relativistic Hedin's equations in which the full photonic degrees of freedom are taken into account. Some further remarks are needed here:
\begin{enumerate}

\item In this framework, the Bethe-Salpeter equations~\eqref{Vertex_Dyson} and~\eqref{Polar_Dyson}  coincide with Eqs. (C.11) and (C.29) already derived in Ref.~\cite{Strinati}.

\item The { \it formal} structure of these non-relativistic QED Hedin's equations is analogous to two cases already studied in the literature:
The first is the relativistic case~\cite{Casalbuoni}, derived by replacing the vector current $j_{\mu}$ with the relativistic charge current $j^{\mu}\equiv\Bar{\Psi}(\vect{x}) \, \gamma^{\mu} \, \Psi(\vect{x})$ (where now $\Psi(\vect{x})$ are spinors and $\gamma^{\mu}$ are the Dirac $\gamma$-matrices). As we have already explained, in this case there is no diamagnetic contribution however.
Finally, in the spin dependent case~\cite{Arya_bierm}, one needs to replace the paramagnetic current with the magnetization operator $m^{i}\equiv\hat{\Psi}^{\dagger}(\vect{x}) \, \sigma^{i} \, \hat{\Psi}(\vect{x})$ (with $\sigma^i$ being the Pauli matrices).

\item The "diamagnetic" term in the photon Green's function $D_{0}$ contributes new diagrams (see Fig.~\ref{Fig1}), similar to the quantum back reaction in scalar QED~\cite{Back-reaction}.

\end{enumerate}
\section{The Exciton-Polariton}\label{polariton_sec}
%
\begin{figure}[b!]
\includegraphics[width=\columnwidth]{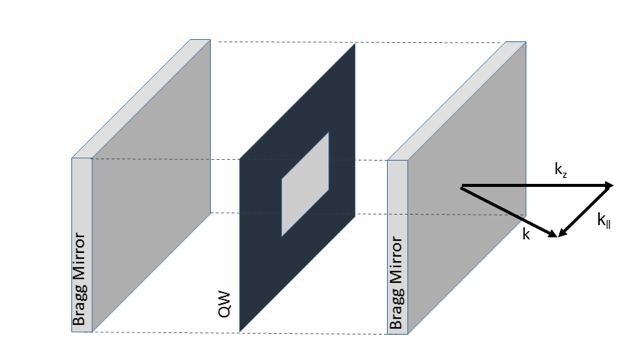}
\caption{\label{Fig2} {\bf semiconductor microcavity Scheme}. The QW is embedded in two Bragg mirrors The wave vector along z direction is quantized whereas the $k_{\parallel}$ is free. }
\end{figure}
In a QW embedded by two Bragg mirrors (microcavity), the wave vector in the $z$ direction perpendicular to the cavity plane is quantized, whereas the in-plane motion is free (see Fig.\ref{Fig2}). This brings us to an important physical phenomenon: the coupling between a planar cavity photon mode and a boson such as an exciton, which leads to a quasi-particle called a polariton. This makes an ideal case in which the generalized Hedin equation can be applied. In order to describe the exciton-polariton, we, first start from the free \emph{transverse} photon Green's function for the photons generated by the external source and trapped in the microcavity. In reciprocal space, Eq.~\eqref{BarePhoton_prop} reads
\begin{equation}
(\mathscr{D}_{(0)})_{ij}(\vect{k},\omega)= \frac{4\pi [\delta_{ij}-(k_{i}k_{j}/k^2)]}{\omega^{2}-c^2k^2+i\eta},
\label{rspD0}
\end{equation}
where $\eta \to 0$ is a regularization parameter. From the Hedin equation ~\eqref{Screened_dy} and by using the expression for $D_{(0)}$ in Eq.~\eqref{rspD0}, the dressed \emph{transverse} photon Green's function is \footnote{we use the identity: $(\omega^{2}-(A+B))^{-1} =(\omega^{2}-A)^{-1} +(\omega^{2}-A)^{-1}B(\omega^{2}-(A+B))^{-1}$}:
\begin{equation}
D_{ij} (\vect{k},\omega)= \frac{4\pi [\delta_{ij}-(k_{i}k_{j}/k^2)]}{\omega^{2} \vect{\epsilon}^{T}(\vect{k},\omega)-c^{2}k ^{2} +i\eta},
\label{rspD2}
\end{equation}
where we have introduced the macroscopic, \emph{transverse} dielectric function $ \epsilon^{T}_{ij}$  \cite{Strinati}:
\begin{equation}
\epsilon^{T}_{ij}(\vect{k},\omega)=\delta_{ij}\left( 1- \frac{\omega^{2}_p}{\omega^{2}}\right)  - \frac{4\pi e^2}{\omega^{2}}\Pi_{ij}(\vect{k},\omega).
\label{kubo}
\end{equation}
Here the plasma frequency $\omega_p=\sqrt{4\pi e^2 \rho /m}$  is obtained from the diamagnetic correction in the bare photon Green's function by keeping only the first order term, see Fig.\ref{Fig1}. The \emph{transverse} polarization $\bf{\Pi}$ can be interpreted as the self energy of the photon in analogy to the standard case. The dispersion relation of the exciton-polaritons are obtained from the poles of $\bf{D}$ in Eq.~\eqref{rspD2} \cite{DelSoleFiorino}:
\begin{equation}
\det\left|\omega^{2}\epsilon^{T}({\bf k},\omega)-c^{2}k^2\right|=0.
\label{disp_rel}
\end{equation}
Considering the $z$ direction as the axis perpendicular to the QW, the photon motion along this direction is quantized as $k_z=\pi M/l_z$, with $M$ being a positive integer and $l_z$ being the thickness of the cavity~\cite{Carusotto}. Therefore, in Eq.~\eqref{disp_rel} $k^2= (\pi M/l_z)^{2}+k^2_{\parallel}$.
\textit{Ab initio} calculations of dispersion relations for excitons have recently been performed  by solving the Bethe Salpeter equation, showing that nowadays they are computationally feasible~\cite{gatti-sottile,cudazzo_disp}.
\section{Conclusions}
\label{conclusion}
In conclusion, we have generalized the Hedin equations in order to take into account the dynamics of the EM transverse field for systems in the non-relativistic regime. In order to do it, we have obtained an effective Hamiltonian in which the degrees of freedom of the internal EM gauge field are integrated out. This leads to a new term coming from the diamagnetic current in the photon Green's function. By using the Schwinger approach, the generalized Hedin equations are derived. The extension proposed here is general in nature. As a possible application, we have focused our attention on the self-consistent solution of these equations in the context of resonant cavities, where the photon-matter coupling is enhanced by photon confinement. In this regard, we have considered one of the most important phenomena taking place in QED microcavities: the exciton-polariton, for which a many body Green function has been obtained. Nevertheless, because our formalism is completely general, it can also be used for \textit{ab initio} studies of polariton-polariton \cite{Luc} and photon-photon effects\cite{Carusotto}. The \textit{ab initio} study of the ultra-strong coupling regime of ISB polaritons will be the focus of future investigations.  

\begin{acknowledgements}
The authors would like to thank G. Vignale for the useful discussions. P.E.T. thanks M. Ladisa and V. Olevano for their critical reading of the manuscript. P.E.T. acknowledges support from  Singapore National Research Foundation (NRF-CRP  8-2011-06  and  NRF2008NRF-CRP002024),
MOE-AcRF   Tier-2   (MOE2010-T2-2-121). M. Milletar\`{i} acknowledges support from the Singapore National Research Foundation (NRF Award No. NRF-NRFF2012-01).
\end{acknowledgements}
%

\appendix
\section{Derivation of the Hedin equations in microcavities} \label{appendix}
In Eq.~\eqref{Heisenberg_eq_mot} the interacting terms (both temporal and spatial terms) are a direct sum in the space-time space. Therefore, they can be addressed separately. Since the temporal component leads to the known Hedin equations, in this Appendix we focus only on the new terms stemming from the external vector field $A_i$. Recalling that the vector source field is $\zeta_i$, we have
\begin{widetext}
\begin{equation}
\hat\xi_{i}(3,3)G_2(1,3;2,3^+)= G_1(1,2)\hat\xi_{i}(3,3) G_1(3,3^+)-\frac{\delta G_1(1,2)}{\delta \zeta_{i}(3)}
\end{equation}
Here the mass operator $M$ is defined as follows:

\begin{equation} \nonumber
\begin{split}
M(1,3) G_1(3,2)&\equiv
\int d3 (D_{(0)})_{ij} (3,1) \frac{\nabla^{i}_{1}}{2m} \hat{\xi}_{j}(3^{+},3^{+}) G_{2}(1,3^{++};2,3^{+})
=\int d3 (D_{(0)})_{ij} (3,1) \frac{\nabla^{i}_{1}}{2m} \left[ G_{1}(1,2)j_{j}(3)- \frac{\delta G_1(1,2)}{\delta \zeta_{j}(3)} \right] \\
&\equiv V^{H}_{i}(1)\frac{\nabla^{i}_{1}}{2m} G_{1}(1,2) -\int d3 \Sigma(1,3)G_{1}(3,2)
\label{Mass_oper}
\end{split}
\end{equation}

where we have introduced the self energy $\Sigma$ and $V^{H}_{\mu}$ is the Hartree term:
\begin{equation}
V^{H}_{i}(1)\equiv \int d2 (D_{(0)})_{ij} (1,2) j_{j}(2)
\end{equation}
With the \textit{total field}: $\Phi_{i} \equiv \zeta_{i} + V^{H}_{i}$, the inverse \textit{dielectric tensor} is
\begin{equation}
\begin{split}
\epsilon^{-1}_{ij}(1,2)\equiv \frac{\delta \Phi_{i}(1)}{\delta \zeta_{j}(2)} &=\delta_{ij}\delta(1,2) +\frac{\delta V^{H}_{i}(1)}{\delta \zeta_{j}(2)}=\delta_{ij} \delta(1,2) +\int d34 \frac{\delta V^{H}_{i}(1)}{\delta j_{k}(3)}\frac{\delta j_{k}(3)}{\delta \Phi_{l}(4)} \frac{\delta \Phi_{l}(4)}{\zeta_{j}(2)}\\
&=\delta_{ij}\delta(1,2) +\int d34 (D_{(0)})_{ik} (1,3)\Pi_{kl}(3,4)\epsilon^{-1}_{lj}(4,2)
\end{split}
\label{diel_inv}
\end{equation}

where we have also defined the polarization $\Pi_{kl}(1,2)\equiv\delta j_{k}(1)/\delta \Phi_{l}(2)$.
The Hedin \textit{photon} Green's function, Eq.~\eqref{Screened_dy}, is obtained by using Eq.~\eqref{diel_inv} and considering that
\begin{align} \label{Dprop}
D_{ij} (1,2) &= \int d3 \epsilon^{-1}_{ik}(1,3)(D_{(0)})_{kj} (3,2) \\ \label{diel_photon_prop}
D_{ij} (1,2)&= (D_{(0)})_{ij}(1,2)+\int d345(D_{(0)})_{ik}(1,5)\Pi_{kl}(5,4)\epsilon^{-1}_{lm}(4,3)(D_{(0)})_{mj}(3,2)
\end{align}
The \textit{irreducible} vertex is defined as
\begin{equation}
\Gamma^{i}(1,2;3) \equiv\frac{\delta G^{-1}_{1}(1,2)}{\delta \Phi_{i}(3)}
\label{irr_vert}
\end{equation}
and with the chain functional derivative, we have
\begin{equation}
\frac{\delta G^{-1}_{1}(1,2)} {\delta a_{j}(3)} =\int d4 \Gamma^{i}(1,2;4)\epsilon^{-1}_{ij}(4,3)
\label{irr_vert_2}
\end{equation}
The electronic self-energy $\Sigma$ Dyson equation~\eqref{SE_Dyson}  is obtained from Eq.~\eqref{Mass_oper} (considering that the dressed photon Green's function is defined in Eq.~\eqref{Dprop}:
\begin{equation}
\Sigma (1,2)\equiv -\imath\int d35 (D_{(0)})_{ij} (3,1) \frac{\nabla^{i}_{1}}{2m} \frac{\delta G_1(1,5)}{\delta \zeta_{j}(3)} G^{-1}_{1}(5,2)
\end{equation}
then by using the identity $\left( \delta G_1/\delta \zeta\right)  = -G_1 \left( \delta G_1^{-1} / \delta \zeta\right)G_1$, we find
\begin{equation}
\Sigma (1,2)\equiv -\imath\int d3456 (D_{(0)})_{ij} (3,1) \frac{\nabla^{i}_{1}}{2m} G_{1}(1,4)\frac{\delta G^{-1}_1(4,6)}{\delta \zeta_{j}(3)}G_1(6,5) G^{-1}_{1}(5,2),
\end{equation}
and we finally arrive at
\begin{equation}
\Sigma (1,2)\equiv -\imath\int d34 (D_{(0)})_{ij} (3,1)\frac{\nabla^{i}_{1}}{2m} G_{1}(1,4)\frac{\delta G^{-1}_1(4,2)}{\delta \zeta_{j}(3)}.
\label{middle_self}
\end{equation}
The Hedin equation \eqref{SE_Dyson} is recovered from Eq.~\eqref{middle_self} by using the \textit{bare} vertex definition equation~(12), Eq.~\eqref{irr_vert_2}, and the definition in Eq.~(\ref{irr_vert}). In this framework, the Hedin vertex equation is the same as that shown in Ref. \cite{Strinati}:
\begin{equation}
\begin{split}
\Gamma^{i}(1,2;3)&= \hat\xi_{i}(3,3')\delta(1,3')\delta(2,3) +\int d4567 \frac{\delta \Sigma(1,2)}{\delta G_1(4,5)} G_1(4,6)G_1(7,5) \Gamma^{i}(6,7;3)=\\
& =\Gamma^{(0)i}(1,2;3) +\int d4567 \frac{\delta
\Sigma(1,2)}{\delta G_1(4,5)} G_1(4,6)G_1(7,5) \Gamma^{i}(6,7;3)
\end{split}
\end{equation}
Similarly, the Dyson equation for the polarization $\Pi_{ij}(1,2)$ is derived, considering the definition $\Pi_{ij}(1,2)\equiv\delta j_{i}(1)/\delta \Phi_{j}(2)$:
\begin{equation}
\Pi_{ij}(1,2)= \hat{\xi}_{i}(1,1')\frac{\delta G_1(1,1')}{\delta \Phi_{j}(2)}= \int d34 \hat{\xi}_{i}(1,1')G_1(1,3) \frac{\delta G^{-1}_1(3,4)}{\delta \Phi_{j}(2)} G_1(4,1')
\end{equation}

\end{widetext}

%

\end{document}